\pdfoutput=1 
\documentclass[cits]{JINST}
                     \let\ifpdf\relax
\usepackage{color,amsmath}
\usepackage[normalem]{ulem}
\usepackage{lineno}
\usepackage{url}
\usepackage{diagbox}
\usepackage{newtxmath,newtxtext}
\usepackage{nicefrac}

\DeclareUnicodeCharacter{2212}{-}
\newenvironment{centermath}
 {\begin{center}$\displaystyle}
 {$\end{center}}

\title{\boldmath Parameterization of Electron Attachment Rate Constants for Impurities in LArTPC Detectors}

\author{Y. Li$^a$~\thanks{Corresponding author: yichen@bnl.gov},
     C. Bromberg$^b$,
     M. Diwan$^a$,
     S. Kettell$^a$,
     S. Martynenko$^a$, 
     X. Qian$^a$,
     V. Paolone$^c$, 
     J. Stewart$^a$,
     C. Thorn$^a$,
     and C. Zhang$^a$,\\
        \llap{$^a$}Physics Department, Brookhaven National Laboratory,
    Upton, NY 11973, USA\\
        \llap{$^b$}Department of Physics and Astronomy, Michigan State University, East Lansing, MI 48824, USA\\
        \llap{$^c$}Department of
Physics and Astronomy, University of Pittsburgh, Pittsburgh, PA 15260, USA
    }

\abstract{The ability of free electrons to drift long distances at high velocities in pure liquid argon under an applied electric field has been exploited for the past forty years to implement detectors with increasingly larger volumes for high energy physics research.  The attachment of free electrons to impurities in the LAr is an important limit on the free instrumented volume of these extremely large detectors, and impurity concentrations as small as 100 ppt can reduce their resolution and efficiency.
In this paper, we summarize the electron attachment rate 
constants as a function of the applied electric field, for common impurities in LArTPCs, 
obtained from data in the literature. We further provide analytical functions to parameterize 
the data, which are useful to compare with new measurements, to model and analyze the performance of existing detectors, and to predict the performance of new detectors. 
}


\begin{document}
\maketitle
\flushbottom
\section{Introduction}
Liquid argon time projection chambers (LArTPC) and calorimetric detectors~\cite{Chen:1976pp, rubbia77, willis74, Nygren:1976fe}
are widely used in neutrino~\cite{Amerio:2004ze,ArgoNeuT2012,Berns:2013usa,Badhrees:2012zz,Acciarri:2016smi,Hahn:2016tia,Abi:2017aow} and dark matter experiments~\cite{DarkSide:2013syn,Badertscher:2013ygt,Zani:2014lea}. 
On the neutrino frontier, the Short-Baseline Neutrino (SBN)~\cite{Antonello:2015lea} program and 
the Deep Underground Neutrino Experiment (DUNE)~\cite{Acciarri:2016crz} will perform a precision search and 
measurement of neutrino oscillations with single-phase multiple LArTPCs, respectively.  
On the dark matter frontier, DarkSide-20k~\cite{DarkSide-20k:2017zyg} will advance the search for direct 
dark matter detection with a dual-phase LArTPC. 

For the detection of the ionization charge in LArTPCs,  
electron attachment to impurities in LAr (such as H$_2$O or O$_2$) is a source of signal attenuation, 
given that the drift velocity of ions is about five orders of magnitude lower than that of 
electrons~\cite{doi:10.1063/1.334552}.  This means that even transient or resonant attachment of an electron to form a negative ion with lifetimes longer than a fraction of the electron drift time will cause loss of signal. 

Attachment to form a stable Ar$^-$ ion has not been reported. Electron attachment to Ar, in the gas phase, has been observed to form excited states at 11.10 eV and 11.28 eV, both with lifetimes of about $10^{-16} $ s~\cite{Hammond:1996}. Since these energies are so far above the mean electron energy at the drift fields sustainable in LAr detectors and the lifetimes are so short compared to drift times, direct attachment to argon atoms does not occur. It remains possible that in pure LAr, other processes, such as attachment to clusters, might occur. However, for practical purposes, the lifetime of free electrons in pure LAr is so long that other fundamental processes, such as diffusion, would limit the ultimate useful drift length in truly pure LAr.

Accurate knowledge and minimization of charge attenuation is critical for LArTPCs, since the collected charge determines the energy reconstruction, and any attenuation will lead to a systematic error in energy determination and a decrease in resolution because of a reduction of signal-to-noise ratio. It may also lead to a loss in efficiency, especially for short or minimum ionizing tracks.  
In a previous publication~\cite{Zhang:2020oaj}, we constructed a model for describing the dynamics of impurity concentrations in LAr detectors and their dependence on materials of construction and operating conditions.  
In this paper, we analyze the relation between impurity concentrations and electron drift lifetimes, characterized by the electron attachment rate constant.

In LAr containing an impurity with an electron affinity (often loosely characterized as an electronegative impurity\footnote{Electronegativity measures the relative attraction of a given atom for the shared electrons when forming a molecule in a chemical bond (see \cite{electronegtivity}). Therefore molecules cannot be assigned electronegativities.  Electron affinity is the binding energy of the electron in a negative ion of an atom or molecule, each in its ground state.}),
the charge $Q$ of the electrons in a swarm decreases exponentially with time as it drifts \cite{bakale1976effect}:
\begin{eqnarray}\label{eq:att}
\label{eq:Q_atten}
\frac{dQ}{dt} &=& -k_A\cdot n\cdot Q, \\
  Q(\tau_D) &=& Q(0) \cdot e^{-k_A \cdot n \cdot \tau_D},
\end{eqnarray}
where $n$ is the concentration\footnote{We express concentrations as mole fractions of impurity in LAr: 1 ppb is 10$^{-9}$ and 1 ppt is 10$^{-12}$.  To convert concentrations in mole fractions to volumetric concentrations multiply by the density of LAr (0.03493 mol/cm$^3$).  One ppt is about the same volumetric density of impurity molecules as 10$^{-6}$ torr of ideal gas at 20$^o$C.} of the impurity, $Q(0)$ is the initial charge of the electron swarm, $Q(\tau_D)$ is the charge after drifting for a time $\tau_D$, and $k_A$ is the electron attachment rate constant for the impurity in LAr. This equation is the basis for most of the measurements discussed in this paper.
The mean lifetime of the charge cluster is 
\begin{eqnarray}
\label{eq:lifetime}
\tau_A=\frac{1}{k_A\cdot n},
\end{eqnarray}
which is referred to as the electron lifetime. The reciprocal of the electron lifetime is the attachment rate.  The attenuation length $\lambda_A$ is defined as:
\begin{eqnarray}
\label{eq:attn_length}
\lambda_A=v_D\cdot \tau_A=\frac{v_D}{k_A\cdot n},
\end{eqnarray}
where $v_D$ is the electron drift velocity in LAr.  The drift velocity as a function of electric field used here was obtained from a global fit to the literature as described in Ref.~\cite{Li:2015rqa}.

Measurements of attachment rates are determined by establishing a uniform electric field between a cathode and an anode, for instance by a pair of parallel plates. The entire apparatus is immersed in the liquid (LAr). Electrons are produced at the cathode by photoelectric emission using a pulsed light source incident on an appropriate photocathode material (usually gold).  The charge of a swarm of electrons leaving the cathode plate, $Q(0)$, will be attenuated during the drift time to the anode where the observed charge is $Q(\tau _D)$.  These charges are converted to voltages by charge sensitive amplifiers connected to the cathode and anode, and then recorded. The drift time is determined from the delay between the cathode and anode signals, and the attachment rate constant is computed for a known concentration and drift distance by inverting Eq.~\ref{eq:att}.  There are many details in the successful implementation of this scheme: grids are added close to the cathode and the anode to screen the slow negative ions produced by attachment so that the charge of electrons only is observed, the drift distance can be made long by the use of field shaping gradient rings, a single charge-sensitive amplifier can be capacitively coupled to both the anode and cathode to ensure identical gains for the two charge signals, etc.  The standard instrument used to implement this procedure at present is the ICARUS Purity Monitor~\cite{Carugno_PM, bettini1991study}.

The electron attachment rate constant $k_A$ can be computed as an integral over energy of the product of the drifting electron energy distribution function and the cross section for the attachment of the electron to an impurity molecule,  both of which are functions of the electron energy. Therefore, $k_A$ is a function of the external electric field.   For electric fields below about 100 V/cm, the electrons are in thermal equilibrium with the liquid, and in this region the electron energy distribution function (EEDF) is independent of electric field.  Therefore the attachment rate is independent of the electric field at low fields.  Also, in the low field limit the drift velocity is proportional to the electric field (i.e. the mobility is constant), so $\lambda_A$ is simply proportional to the field. There is more discussion of the physics of the attachment process in Section~\ref{Sec:Calculations}, where we describe a simple approximate procedure for computing attachment rate constants, and compare the results to the data.

In this paper, we summarize measured electron attachment rate constants reported in the literature for six impurities in LAr as a function of the applied electric field. We further provide analytical functions that parameterize this data, which can be used to compare with new measurements and to interpolate between measured points and to a limited extent, to extrapolate beyond the measured ranges of electric field values. Interpolation can be particularly useful for comparing lifetimes at one operating field to those at another field. The functional forms can also be useful for modeling detector performance and in comparing energy reconstruction from different LArTPCs.

\section{Global Fit to existing Data}

\subsection{Data from the literature}
 The data sources for the several impurities that we found in the literature are listed in Table~\ref{table:data}, along with the quantities measured, how concentrations were determined, and the thermal conditions of the LAr. The data were extracted from bitmap images of the appropriate graphs in the PDF files obtained from the journal websites for each of the papers.  Each bitmap image was imported into a drawing program (CorelDraw).  By enlarging the image, it was straightforward to  make an accurate estimate of the centroid of the pixels comprising each data point symbol.  The coordinates of the vertices of the lines segments representing the data, error bars, and axes were digitized and scaled to the domain and range on the original axes as indicated in the respective publications. 

\begin{table}[htp]
\centering
\caption{Data for the global fit of attachment rate}
\label{table:data}
\begin{tabular}{|l|l|l|l|l|l|}
\hline
Data Set                & Impurities  &  Data Type & $n$ from & Thermal Conditions & Ref \\
\hline
Adams {\it et al.}  & O$_2$ & $k_A$ & volume & Measurements at 5 Temp & \cite{adams2005purity}\\
Biller {\it et al.}     & O$_2$, N$_2$      & $\lambda_A^{-1}$ & volume &  LN$_2$, no T,  P $\approx$ 1 atm &\cite{Biller:1989yq}\\ 
Bakale {\it et al.} & O$_2$, N$_2$O, SF$_6$ & $k_A$ & volume & LN$_2$, 800<P<1000 Torr &\cite{bakale1976effect}\\
Barabash {\it et al.}  & N$_2$ & $v_D$ \& $Q/Q_0$ & volume & P=6.9 bar, 105<T<109K & \cite{barabash1984}\\
Bettini {\it et al.}  & O$_2$, CO$_2$  & $\tau_A$  & volume & LAr at atm $\Rightarrow$ 87.3K &~\cite{bettini1991study}\\
Carls {\it et al.} & H$_2$O & $\tau_A$ &  liquid & no T, P$\sim$ 3 psig  & \cite{microboone2016measurement} \\
Hofmann {\it et al.} & O$_2$, N$_2$ & $\lambda_A$   & gas / H$_{xx}$ &       LN$_2$, no T, P $\approx$ 1 atm  & \cite{hofmann1976production}\\ 
Sakai {\it et al.} & N$_2$ & $\tau_A$   & volume &   80K, 84K, 88K all $\pm0.5$K  & \cite{hofmann1976production}\\ 
\hline
\end{tabular}
\end{table}

The attachment rate constants $k_A$ are reported explicitly only by Bakale {\it et al.}~\cite{bakale1976effect} and by Adams {\it et al.}~\cite{adams2005purity}. For the remainder of the literature the attachment rate constants are derived from the data provided.  In Bettini {\it et al.}~\cite{bettini1991study} and Hofmann {\it et al.}~\cite{hofmann1976production} electron lifetimes $\tau_A$ and attenuation lengths $\lambda_A$, respectively, are reported as a function of electric field $E$ for several concentrations.  These can be converted to attachment rate constants using an inversion of Eq.~\ref{eq:lifetime}.  Biller {\it et al.}~\cite{Biller:1989yq} reports inverse attenuation lengths, $\lambda_A^{-1}$, as a function of added impurity concentration for several fields.  Each of these data sets was fitted with a straight line to extract the slope ${d{\lambda_A}^{-1}}/{dn}$ which,  from Eq.~\ref{eq:attn_length}, gives the attachment rate constant divided by the drift velocity.  Barbash {\it et al.}~\cite{barabash1984} give plots of drift velocity and charge attenuation over the drift distance (3 mm) for five fields and three nitrogen concentrations. 
The nitrogen concentrations are so large (7\%, 16\% and 25\%) that the drift velocity they report is significantly different than that of pure argon.  The attachment coefficient was computed using Eq.~\ref{eq:Q_atten} and the inverse of Eq.~\ref{eq:attn_length} with the appropriate drift velocity for each concentration.  Sakai {\it et al.}~\cite{Sakai1984} report $\tau_A$ as a function of field for three concentrations.  A portion of this data was measured at three temperatures, 80K, 84K and 88K, all $\pm0.5$K, and showed no changes with temperature.  Therefore, in the analysis of this data we have grouped all measurements together, regardless of the temperature.

The numerical values for the attachment coefficients for all of the data extracted from the literature can be found in the Appendix.  These data are plotted as colored points with error bars in Fig.~\ref{fig:attachment_rates} for SF$_6$, N$_2$O, O$_2$,CO$_2$, and H$_2$O; and in Fig. \ref{fig:attachment_rates_N2} for three sources of data for N$_2$.  For the data of Bakale {\it et al.} and Sakai {\it et al.}, which have no uncertainties specified for each point, we have assigned a constant fractional error to each data point, with a value chosen to make $\chi^2/\text{DOF}=1$.  The values of these assigned fractional uncertainties are given in the appropriate tables in the Appendix.

Concentrations of the impurities were determined by various methods, as indicated in column 4 of Table~\ref{table:data}.  The designation "liquid" means that a sample of the liquid was vaporized and analyzed by commercially available instruments to determine the relevant concentration.  The designation "volume" means that a known volume of the impurity gas was introduced into the cryostat containing a known volume of LAr.  The assumption was made that all of the impurity entered the liquid.  In equilibrium, an impurity will distribute between the gas and liquid in the ratio of the impurity concentrations
\begin{eqnarray}
\label{eq:Henry}
H_{xx}=\frac{c_\text{X, GAr}}{c_\text{X, LAr}},
\end{eqnarray}
which is known as Henry's coefficient.  Since concentrations are expressed as mole fractions (moles of impurity per mole of LAr) the attachment rate constants have the units of s$^{-1}$ and Henry's coefficient is dimensionless.  For a detailed discussion on the role of Henry's coefficient in the distribution of impurities in LAr see~\cite{Zhang:2020oaj}.  The conditions under which equilibrium can be expected are also discussed there, as summarized in the following paragraph.

There are two principal requirements in ensuring that gas and liquid volumes maintain an equilibrium state.  If the liquid evaporates very rapidly, the contact time at the surface may be insufficient to allow exchange of molecules to establish equilibrium.  For example, it is certainly possible to boil the liquid so rapidly (to “flash” it to vapor) that no equilibrium can be established. However, as discussed in Ref.~\cite{Zhang:2020oaj} section 2.1.1, at cryogenic temperatures the evaporation must be very rapid since the mass accommodation coefficient (the analog of the “sticking” factor for adsorption on a surface) at cryogenic temperatures is very large, especially for substances that are in a condensed state at these temperatures. Ref.~\cite{Zhang:2020oaj} demonstrated that even with rapid boiling in a small system it was difficult to break equilibrium.  However, even if equilibrium conditions exist at the surface, it is possible that the bulk of the volumes fail to reach equilibrium since mixing of the material at the surface into the bulk may be insufficient. Diffusion is too small in liquids to homogenize even volumes of a few liters  and only thermal convection (which can be small with  little thermal input) or mechanical agitation can ensure adequate mixing. This is clearly demonstrated in Ref.~\cite{Li:2015rqa}.  Between the limits of no evaporation and “flash” evaporation there is a wide range of boiling rates that will establish essentially equilibrium conditions in closed systems~\cite{Zhang:2020oaj}.  From the (often quite brief) descriptions of the procedures for the data  quoted here, it seems reasonable that the gas and the liquid were in equilibrium when the measurements were made.

For the six cases labeled "volume" in the table, the assumption that Henry's coefficient is zero (i.e. that all of the impurity injected into the system is in the LAr) leads to a very small error.  An expression for the actual final concentration in the liquid, after injection of $n_{X \text{, Add}}$ moles of impurity and the subsequent equilibration with the gas, can be obtained by solving a set of equations involving the impurity ($X$) concentrations $c_X$, the phase volumes $V$, and the argon molar phase densities $\rho_\text{M}$:
\begin{equation}
   c_{X\text{,LAr,Final}} = c_{X\text{,LAr,Entire}}\;(1\;+\;H_\text{xx}\;\alpha_\rho\;r_\text{V})^{-1}.
    \label{eq_conc_err}
\end{equation}
where $c_{X\text{,LAr,Entire}}\;=\;n_{X \text{, Add}}/n_{\text{ LAr}}$ is the impurity concentration in the LAr that would be obtained if all of the injected impurity were in the liquid, $\alpha_\rho=\rho_\text{M,GAr}/\rho_\text{M,LAr}$ is the ratio of the densities of GAr  to LAr, and $r_\text{V}=V_\text{Gas}/V_\text{Liq}$ is the ratio of the gas volume to the liquid volume.
Since the molar density ratio of LAr to GAr is 242 at the normal boiling point, unless the gas volume is very much larger than the liquid volume, Henry's coefficient is inconsequential and essentially all of the impurity is in the liquid.  The ratio of the liquid and gas volumes are not much larger than one for all the measurements discussed here, and Henry's coefficients for all the relevant gases are not very large (see Table~\ref{table:properties}).  Therefore, for all of the measurement quoted here, the largest error from this source occurs for $N_2$ in Biller {\it et al.}, where the gas to liquid volume ratio is $\sim$2, resulting in an error of $\sim$2\%.  In the measurements of Hofmann {\it et al.} the concentration in the gas is directly measured, and this value is converted to that in the liquid by dividing by Henry's coefficient for O$_2$. In the measurement of Carls {\it et al.} the concentration in the liquid was determined by converting the liquid to gas in a tube terminating in the liquid and measuring the concentration in the resulting gas.

\begin{table}[htp]
\centering
\caption{Properties relevant to the determination of attachment rates}
\label{table:properties}
\begin{tabular}{|l|l|l|l|l|l|}
\hline
Species     & $E_A\;(eV)$  &  $\Delta E_\text{diss}\;(eV)$ & $H_{xx}$ & Solubility \\
\hline
SF$_6$ & 1.07 &	3.38 &	(1.1$\times 10^{-5}$) &	unknown\\
O$_2$ &	0.451 &	5.12 &	0.91 &	1\\
N$_2$O &	0.22 &	2.71 &	(1.7$\times10^{-6}$) &	0.000075\\
H$_2$O &	$\sim$0 &	5.15 &	(4.1$\times10^{-3}$) &	$<10^{-8}$\\
CO$_2$ &	-0.6 &	5.45 &	(1.6$\times10^{-4}$) &	$2.8\times 10^{-6}$\\
N$_2$ &	-1.6 &	9.76 &	3.5 &	1\\
\hline
\end{tabular}
\end{table}

We found two papers that provide lifetime and concentration measurements for water in LAr. The first of these, Andrews {\it et al.}~\cite{andrews2009system}, reports measured water concentrations in the gas.  For these measurements, the conditions of the sampled gas appear to be far from equilibrium with the liquid. It is therefore difficult to relate, with confidence, the measured gas concentrations to those in the liquid, so we did not use this data.  However, in the second paper, Carls {\it et al.}~\cite{microboone2016measurement}, which gives H$_2$O and O$_2$ concentration data taken after the initial filling of the MicroBooNE cryostat, LAr is sampled with a tube dipping into the liquid. The liquid is converted to gas as it travels up the tube, and the concentration of water in this gas is measured as described in the paper. If the evaporation is done in the  ``sudden" (thermodynamically adiabatic\footnote{For the design of an adiabatic liquid sampling system see~\cite{preston1971solubilities}.}) limit, then the gas concentration is that of the liquid. Figure 2 in Ref.~\cite{andrews2009system} displays the concentrations of O$_2$ and H$_2$O in the LAr during a period of 18 days following the start of re-circulation of the LAr through a purifier. Figure 6 in Ref.~\cite{andrews2009system} is a plot of the charge attenuation over the drift distance of a long ICARUS purity monitor (50 cm).  This graph also marks the charge attenuation ratios corresponding to 3 ms, 6 ms, and 9 ms. We have used these three points to establish a logarithmic scale, as required by Eq.~\ref{eq:att}, to convert the charge ratios to lifetimes.  There are 25 points in this lifetime data set, during the interval between 13 an 18 days, in common with the concentration data set.  For the time of each point in the concentration data set we have obtained the corresponding lifetime, and for each pair of  values we computed the attachment rate constant, using Eq.~\ref{eq:lifetime}. The mean and standard deviation of these values gives an attachment rate constant for water in LAr of $(7.80 \pm 1.5) \times 10^{10}$ s$^{-1}$ at an electric field of 32 V/cm.  This electric field is not stated in Ref.~\cite{andrews2009system}; we have deduced it from the stated drift time and length using the value of the electron mobility in LAr~\cite{Li:2015rqa}.

Two processes can cause significant systematic uncertainties in the above evaluation of the attachment rate constant from the MicroBooNE data set~\cite{microboone2016measurement}.  The first arises from the facts that the purity monitor and the sampling point are at some distance from each other and the that the water concentration may not be uniformly distributed throughout the entire volume. We assign $\pm 20\%$ for this uncertainty.  The second is that O$_2$ may contribute to the measured lifetime.  The O$_2$ concentration measurement fell below the detection threshold of 100 ppt after 6.5 days and remains there for the remainder of the measurement.  We therefore estimate a residual concentration less than 10 ppt of O$_2$; removing this would systematically lower the attachment rate constant of water by 10\%.  The sum in quadrature of these two systematic uncertainties is represented in the error bar for the data point for attachment to water shown in Fig.~\ref{fig:attachment_rates}.

A third systematic uncertainty arises from the possibility that the evaporation occurred under conditions at or near equilibrium between the liquid and gas (as is the case if the two liquids remain in good thermal contact and the evaporation rate per contact area is small), in which case the concentration measured in the gas will be Henry's coefficient times the concentration in the liquid, and the above attachment rate constant must be multiplied by this factor. 

Unfortunately, the value of Henry's coefficient for H$_\text{2}$O in LAr has not been measured.  However, an estimated value is reported by the NIST program REFPROP v10, using an EOS model fitted to a large body of data; this value is $4.1\times 10^{-3}$ at 90 K~\cite{nistrefprop}.
If this estimate for Henry's coefficient were correct, then would we find an attachment rate constant of $3.2 \times 10^{8}$ s$^{-1}$.  This should be considered as a lower limit on the true value consistent with this data, since it is likely that the sampling was in fact more nearly adiabatic than equilibrium,and the true value is probably much closer to the larger value quoted above.  To represent this uncertainty, we do not include it in the stated value but indicate it in Fig.~\ref{fig:attachment_rates} as an arrow pointing to the lower limit.

Of all the sources of the data used here, only Adams {\it et al.} state the temperature of the LAr at which the measurements were made; the others give some (or very little) information from which the temperature might be inferred.  Notably, Adams {\it et al.} report measurements of $k_A$ at five temperatures from 88.6K to 91.9K and for six fields.  From this data we deduce a mean temperature coefficient for attachment to O$_2$ of $k_A^{-1}\;d\,k_A/dT=0.06\pm0.01$ K$^{-1}$, which is quite independent of field over the range they report (2.2 to 6.2 kV/cm).  Their values for $k_A$ as quoted here have been extrapolated to 87.3K. This was done in order to match the data of Bettini {\it et al.}, who used a LAr bath at atmospheric pressure to contain the LAr measurement cell, establishing a measurement temperature of 87.3K.  Bakale {\it et al.} state that the pressure in the measurement cell was between 800 and 1000 Torr, which implies a LAr temperature between 87.8K and 90.0K.  The data from Carls {\it et al.} was measured in MicroBooNE, which typically operates at about 3 psig, and this implies a temperature of 89K.  For the measurements taken from the remaining two sources, we only know that the LAr was cooled by a bath of LN$_2$, apparently with weak thermal coupling, and presumably at atmospheric pressure. We assume that the LAr was maintained on the boiling line at atmospheric pressure,  so a temperature near 87.3K.  Because of these uncertainties we have not corrected any of these three data sets for temperature differences.
The temperature variation within and between the various measurements contributes 5\% to 10\% systematic error.  The systematic errors in concentration, when reported, are between 10\% and 20\%  which dominate the total systematic error in the attachment coefficients discussed here.

\subsection{Fitting Functions}
The attachment rate data is fitted with a constrained rational polynomial:
\begin{equation}
    k_A=10^p\cdot\frac{\frac{a_1}{b_1}+a_1 E+a_2 E^2+ a_3 E^3 + a_4 E^4}{1+b_1 E+ b_2 E^2+b_3 E^3+b_4 E^4},
    \label{eq_pade}
\end{equation}
where $E$ is the electric field in units of kV/cm and $k_A$ is the attachment rate constant in units of $s^{-1}$, $p$ is a scaling factor introduced to keep the values being fitted close to unity.  The term $\frac{a_1}{b_1}$ in the numerator ensures that
$\lim\limits_{E\rightarrow0} \frac{dk_A}{dE}=0$.  This constraint is imposed to force the fit function to be constant at low field, as attachment rate constants must be. 

We have chosen to use rational polynomial functions, rather than simple polynomial functions, to fit the attachment data, motivated by the unique capabilities of Pad\'e approximants in optimally representing mathematical functions~\cite{Press1989numerical}. In general, compared to simple polynomials, rational polynomials
  1) can produce a larger variety of shapes;
  2) are less oscillatory, providing more credible interpolation between data points;
  3) can easily be designed to conform to {\it a priori} behavior outside the domain of the data, providing more useful extrapolation;
  4) can model complicated behavior with fewer terms (the total of numerator and denominator polynomial terms) than a simple polynomial with the same number of terms, allowing more degrees of freedom for a given data set.

For our purposes, the potential disadvantage of using rational polynomials to fit this data is that the best fit function can result in a pole (a root in the denominator) between two adjacent data points, making interpolation in the region between these two points useless.  To avoid this problem, for each data set we have found the best fit for each of the rational polynomials with a numerator and a denominator order from 1 to 4 inclusive (16 rational polynomials), and then selected the one that has the lowest $\chi^2$ per degree of freedom ($\chi^2/\text{DOF}$) from all those for which the denominator has no real roots in the range from zero to 100 kV/cm.
The coefficients of the best-fit rational polynomial obtained in this way for each impurity are listed in Table~\ref{table:table_fit}. There is no fit for the water data, since there is only one point.  

\begin{table}[!htbp]
\caption{Parameters for the best fit to data of Table~\ref{table:data} using Eq. \ref{eq_pade}.}
\label{table:table_fit}
\centering
\begin{tabular}{|c|c|c|c|c|c|c|}
 \hline
Param. & SF$_6$  &  N$_2$O & O$_2$ & CO$_2$  \\
 \hline
p & 16 & 14 & 11 & 11  \\
a$_1$ & 86.5006 & 0.0369405 & 76.2749 & 2.14817  \\
a$_2$ & 351.26 & 0.585551 & 4.24596 & 11.9613 \\
a$_3$ & 13.8193 & -0.00183436 & 0 & 0 \\
b$_1$ & 52.8982 & 1.76938 & 1.88083 & 1.22992  \\
b$_2$ & 547.61 & 0.183879 & 2.62643 & 0   \\
b$_3$ & 100.67 & 0 & 0.0632332 & 0  \\
b$_4$ & 0 & 0 & -0.000211009 & 0  \\
$\chi^2$/DOF & 1.00 & 1.00 & 1.04 & 0.0052 \\
 \hline
\end{tabular}
\end{table}

\begin{table}[!htbp]
\centering
\begin{tabular}{|c|c|c|c|c|c|c|c|}
 \hline
Param. & N$_2$(Bil+Hof) & N$_2$(Barabash) & N$_2$(Sakai) \\
 \hline
p &  10 & 6  & 6 \\
a$_1$ & 2.94322$\times 10^6$ & -0.0506859 & 6.03117$\times 10^7$ \\
a$_2$ & -501897 & -0.0295819 & -947289 \\
b$_1$ & 8.53556$\times 10^6$ & -0.0869897 & 1.3608$\times 10^6$  \\
$\chi^2$/DOF & 9.8 & 0.0065 & 1.00  \\
 \hline
\end{tabular}
\end{table}

\begin{figure}[!htbp]
\centering
\includegraphics[angle=0,width=5.75in]{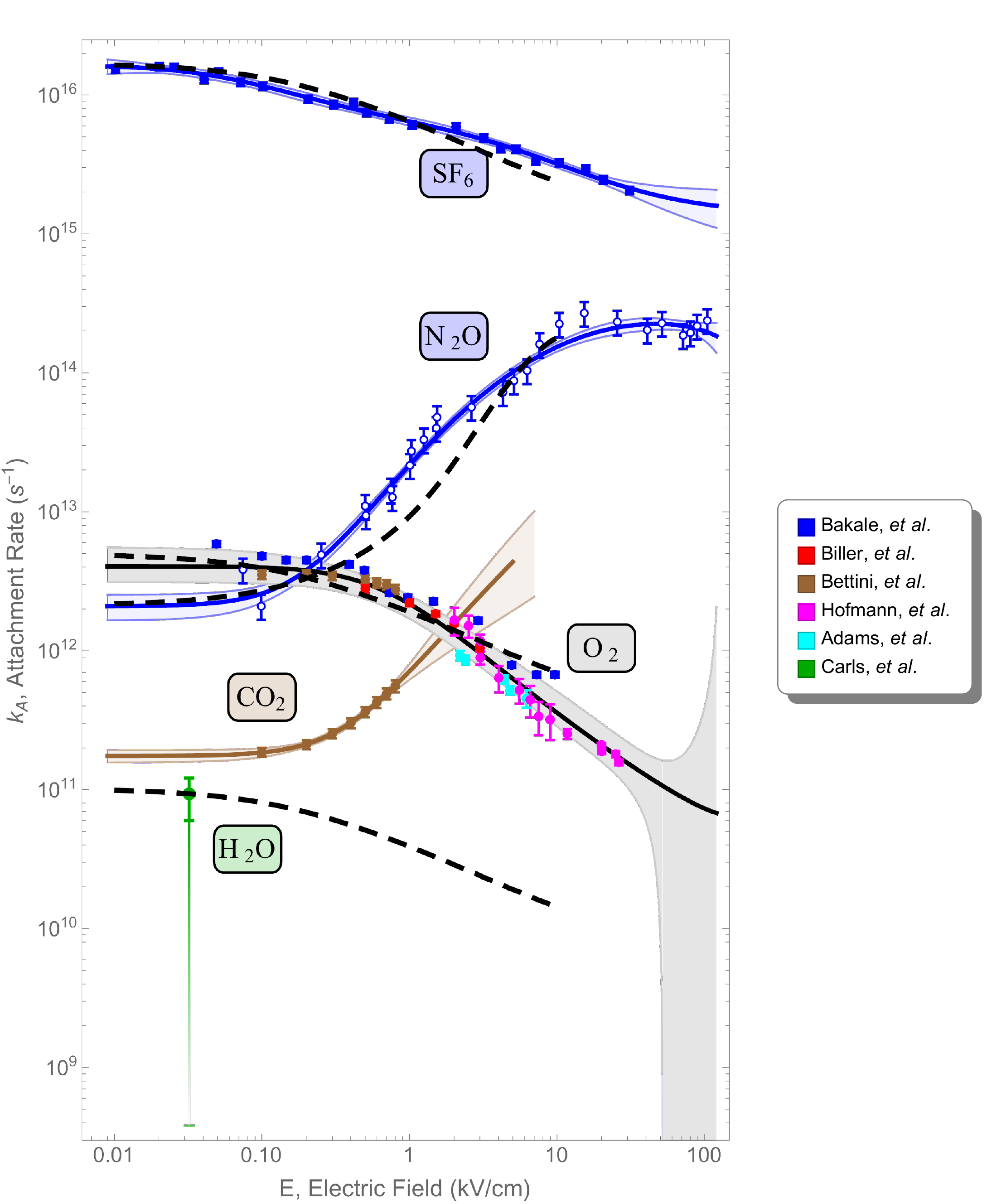}
\caption{The electron attachment rate constant as a function of the applied electric field for five impurities in LAr.  The heavy solid lines are the best-fit curves using the parameters in Table~\ref{table:table_fit}, and the light lines with shading between them indicate the 68\% confidence uncertainty bands for the fit.  Open circles are used for the N$_2$ Bakale {\it et al.} data to distinguish them from the nearby O$_2$ data.  The citation for each source of data in the legend is given in Table~\ref{table:data}.  The numerical values for all data points can be found in the Appendix. The heavy black dashed lines are the result of a calculation of the radiative attachment rate as described in Section~\ref{Sec:Calculations}.}
\label{fig:attachment_rates}
\end{figure}

\begin{figure}[!htbp]
\centering
\includegraphics[angle=0,width=5.9in]{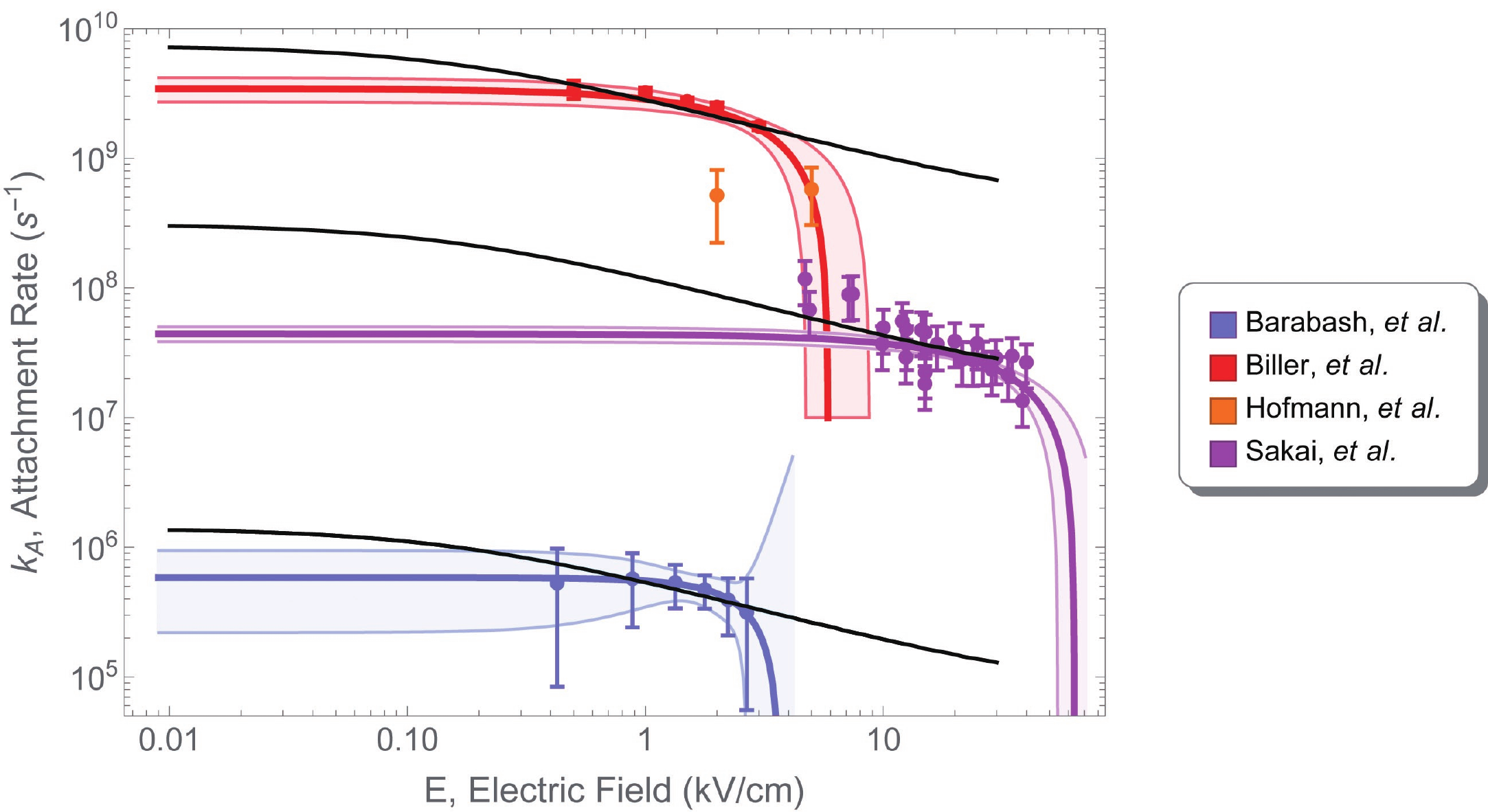}
\caption{The electron attachment rate constants for N$_2$ in LAr as a function of the applied electric field from four sources.  The heavy colored lines are the best-fit curves using the parameters in Table~\ref{table:table_fit}, and the light colored lines with shading between them indicate the 68\% confidence uncertainty bands for the fit.  The black lines are the best fit function for the O$_2$ attachment rate constant (shown in Fig~\ref{fig:attachment_rates}), but scaled by factors of 1.5$\times10^{-3}$, 6.5$\times10^{-4}$, 1.5$\times10^{-4}$ and 2.8$\times$10$^{-7}$ reading from top to bottom, to best represent each of the four N$_2$ data sets.}
\label{fig:attachment_rates_N2}
\end{figure}

The values of the best fit rational polynomial as a function of electric field for each impurity are shown by the thick black solid lines in Fig.~\ref{fig:attachment_rates} and~\ref{fig:attachment_rates_N2}.  The uncertainty bands for each best-fit rational polynomial, at the 67\% confidence level, are shown as the thin colored lines with filling between the upper and lower uncertainty values.  The dashed black lines are the result of a simple calculation of the attachment coefficient from the attachment cross section, described in the next section.

There are five independent determinations of the attachment rate constant for oxygen (see Table~\ref{table:data}), providing a total of 44 data points. The best-fit polynomial for this entire set has a $\chi^2/\text{DOF}$ value of 12.1.  This large value is a consequence of the fact that the individual uncertainties assigned to each point are statistical only and do not represent the systematic uncertainties of each set.  A close comparison of the sets reveals these systematic inconsistencies: for instance, the Bakale {\it et al.} data is systematically higher than the other sets at both low and high fields.  To estimate, and correct for, the systematic uncertainties we have produced fits to all combinations of 2 through 5 of all the sets.  As well as fitting the data as is, we have also allowed the scales of individual sets to vary to obtain an overall best fit.  To constrain the overall scale, the Adams {\it et al.} and Bettini {\it et al.} data sets are always included with no adjustment, since they span the range of the entire set and for these two sets the temperature is accurately known.  This yields a set of 91 best-fit functions.  To obtain the best overall fit we have averaged the values for this set of functions, and re-fitted this "pseudo-data" to a single best rational polynomial.  This produces the parameters shown in Table~\ref{table:table_fit} and the solid line for O$_2$ in Fig.~\ref{fig:attachment_rates}.  To indicate the systematic uncertainty, we have produced upper and lower bounds from the set by dropping the lowest and highest 15.9\% of the values at each field, and then fitting the highest and lowest remaining values at each field to determine the 68\% uncertainty band.  This band of values is also shown in Fig.~\ref{fig:attachment_rates} and~\ref{fig:attachment_rates_N2}.  The $\chi^2/\text{DOF}$ value for the entire O$_2$ data set, ignoring the quoted statistical errors and computed instead using these error limits, is 1.04.  

\subsection{Calculation of Attachment Rate Constants}
\label{Sec:Calculations}
Attachment of an electron to a molecule can occur via several mechanism, including direct (radiative) attachment, in which an excited compound electron-molecule system decays radiatively to a ground state; dissociative attachment, in which the compound state decays by breakup leaving an excited fragment and the electron attached to another fragment of the molecule; and three body collisions, in which a third body participates in the reaction.  The reaction rate for an electron and a molecule X moving toward each other at a speed $v$ to produce a product Y is proportional to the cross section $\sigma_{e+X\rightarrow Y}(v)$,

\begin{equation}
  \frac{dn_e}{dt} = -v \,\sigma_{e+X\rightarrow Y}(v) \, \,n_e \,n_X,
\label{eq_basic}
\end{equation}
so that for a given speed the reaction rate per electron per attaching molecule is $k_{e,X,Y} = v \,\sigma_{e,X,Y}(v)$.  For a probability density distribution of speeds $\rho_v(v)$, the total rate to produce Y is the integral over all speeds
\begin{equation}
  k_{e,X,Y} = \int_0^{\infty} v \, \sigma_{e+X\rightarrow Y}(v) \, \rho_v(v) \, dv.
\label{eq_rate_v}
\end{equation}
The total rate of loss of electrons is the sum of the attachment rates over all final channels Y.  Since cross sections are generally reported as a function of energy, it is convenient to transform the expression for the rate constant into an integral  over energy.  Using $\epsilon = \nicefrac{1}{2} \; m_e v^2$ for the energy, we find
\begin{equation}
  k_{e,X,Y}(E) = (2/m_e)^{1/2} \int_0^{\infty} \sqrt{\epsilon} \, \sigma_{e+X\rightarrow Y} \, \rho_{\epsilon}(\epsilon,E) \, d\epsilon,
\label{eq_rate_E}
\end{equation}
where $\rho_\epsilon = \rho_v ( \sqrt{2} \, \epsilon/m_e)$ is the electron energy distribution function (EEDF) which is a function of the electron energy and the electric field applied to the LAr. The reduced mass of the colliding electron molecule pair has been replaced with the electron mass.

The EEDF is obtained by solving the Boltzmann transport equation.  For gasses this is accomplished by the code Magboltz~\cite{Biagi1999}, using the measured cross sections for electron scattering on argon (or a number of other molecules).  For liquids, coherent scattering on density fluctuations in the LAr must also be included.  This has been done by Atzrashev and Timoskin~\cite{Atrazhev}, but that code is not available to us.  Instead of obtaining the EEDF by solving the Boltzmann equation, we propose as a simple alternative the following ansatz: we take the EEDF to be the Maxwell distribution function in energy,

\begin{equation}
  \rho_{M}(\epsilon,T)\, d\epsilon = \left(\frac{4}{\pi}\right)^{1/2}\, \left(\frac{1}{k\,T}\right)^{3/2}\, \sqrt{\epsilon}\, e^{-\epsilon / k\,T}\,d\epsilon,
\label{EEDF}
\end{equation}
in which $k$ is the Boltzmann constant, $k\,T$ is the electron temperature in energy units.   For this temperature we use the transverse diffusion coefficient, expressed as an energy via the Einstein-Smoluchowski relation~\cite{einstein1905uber}
\begin{equation}
  kT(E) = \frac {e\,D_T(E)} {\mu},
\label{Einstein}
\end{equation}
where $D_T(E)$ is the transverse diffusion coefficient in an electric field E, $\mu$ is the electron mobility, and $e$ is the electron charge. Our motivation for this procedure is that the Maxwell function properly describes the distribution of velocities of identical particles in thermal equilibrium and interacting by instantaneous elastic collisions in the absence of an applied field \cite{morse1971} - perhaps a reasonable first approximation for argon at zero electric field.  The diffusion "temperature" then introduces an empirical, and hopefully utilitarian, estimate of the departure from the zero field condition. The utility of our crude estimate will be seen.

With this modified Maxwell electron energy distribution, the total rate constant for electron attachment to X is computed as
\begin{equation}
  k_{e,X,Y}(E) = \frac{1}{\sqrt{\pi \, m_e}}\, \left(\frac{2}{kT(E)}\right)^{3/2}\, \int_0^{\infty} \epsilon\, \sigma_{e+X\rightarrow Y} \, e^{-\epsilon / kT(E)}\,d\epsilon.
\label{eq:attach_form}
\end{equation}

\begin{figure}[ht]
\centering
\includegraphics[angle=0,width=4.5in]{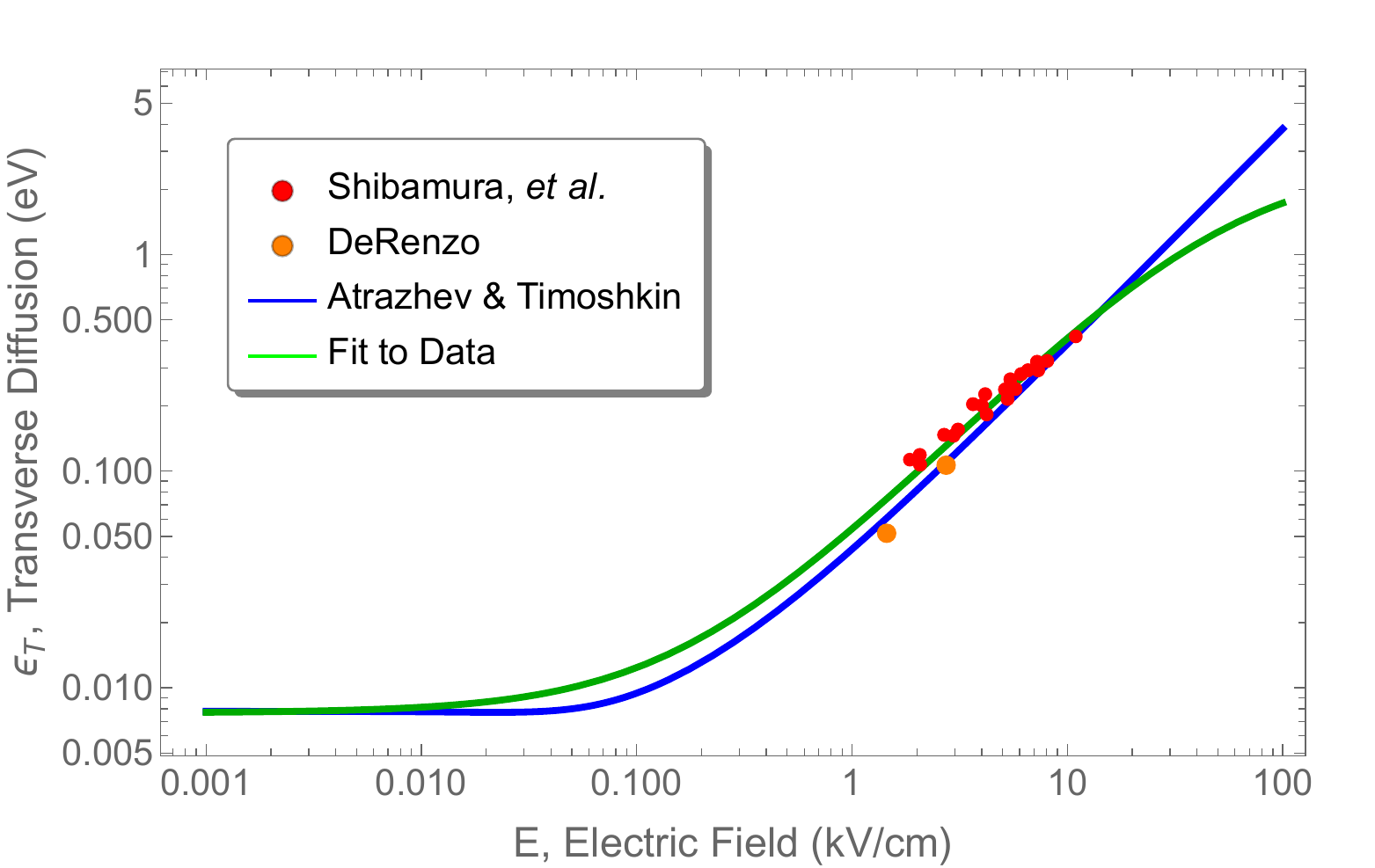}
\caption{Transverse electron diffusion temperature of electrons in LAr, showing measurements, and best fit rational polynomials to the data and to the calculation reported by Atrazhev and Timoshkin.}
\label{fig:trans_diff}
\end{figure}

The transverse diffusion coefficient for electrons in LAr has been measured by Shibamura {\it{et~al.}}~\cite{Shibamura} and by DeRenzo~{\it{et~al.}}\cite{Derenzo:1974ji}, and it has been calculated by Atrazhev and Timoshkin~\cite{Atrazhev}.  Figure~\ref{fig:trans_diff} shows the results of a fit of rational polynomials to these data and to the calculation.  Both give similar results when used to calculate attachment coefficients for fields below 30 kV/cm.  Since it better represents the measurements, we use the function fitted to the data in calculations and restrict calculations to electric fields less than 40 kV/cm.

The first step of attachment to a multi-atomic molecule AB is the formation of a compound excited state
\begin{equation}
 \text{AB}\,+\,e^-\,\rightarrow\,\text{AB}^{*-}.
\label{eq:compound}
\end{equation}
The compound state AB$^{*-}$ then decays into one of three channels
\begin{align}
\label{eq:auto_diss}
 &\text{AB}^{*-}\,\rightarrow\,\text{AB}\,+\,e^-,\\
\label{eq:direct_att}
 &\text{AB}^{*-}\,\rightarrow\,\text{AB}^-\,+\,\gamma,\\
\label{eq:dissoc_att}
 &\text{AB}^{*-}\, \rightarrow \,\text{A}^-\,+\,\text{B}\;\text{ or }A\,+\,B^-.
\end{align}
Reaction~\ref{eq:auto_diss} is auto-detachment.  The lifetime of this decay varies over a wide range, from less than a vibrational period (10$^{-14}$ s) to a few ms if the transition is hindered by the poor overlap of the wavefunctions of the compound state and the ground state.  Reaction~\ref{eq:direct_att} is radiative decay to the ground state, which can only occur for molecules with a positive electron affinity, with a typical lifetime of 10$^{-14}$ s.   Reaction~\ref{eq:dissoc_att} represents dissociative fragmentation into a stable negative and a neutral fragment, with a typical lifetime of less than 10$^{-12}$ s.
For the molecules discussed here, dissociative attachment is relatively unimportant below 40 kV/cm here (except for N$_2$O), since their dissociation energies are above 3 eV, which corresponds to a field greater than $\sim50$ kV/cm (see Fig.~\ref{fig:trans_diff}).  Therefore we include dissociative attachment only in calculating attachment rates for N$_2$O. 

Radiative attachment is a compound state problem which proceeds through an excited combined electron-molecule state AB$^{*-}$, for which the cross section averaged over all resonances is given by Hauser-Feschbach theory~\cite{hauser1952inelastic}. This cross section is the product of a statistical factor $F_\text{stat}$ and the square of the de Broglie wavelength of the electron  $\lambda=h/p$, where $h$ is the Planck constant, $p$ is the momentum of the electron, which is inversely proportional to electron energy:
\begin{equation}
 \sigma_F(\epsilon) = F_\text{stat}\;\frac {\lambda^2} {4 \pi} =  F_\text{stat}\;\frac{h^2}{8\pi\,m_e\,\epsilon}.
\label{eq:max_sigma}
\end{equation}
Measurements of the attachment cross section at low energies not are available for any of the molecules discussed here, except for SF$_6$.  The total electron attachment cross section for SF$_6$ has been measured by Christophorou~\cite{Christophorou} and, as expected for direct attachment through a compound Feshbach resonance, the data shows a roughly $\epsilon^{-1}$ dependence below the lowest dissociation threshold (SF$_5\rightarrow$SF$_5$ + F$^-$).  We use the cross section of Eq.~\ref{eq:max_sigma} to evaluate Eq. \ref{eq:attach_form} for the attachment rate to SF$_6$.  The best representation of the data is obtained with $F_\text{stat}=0.36$.
We have similarly scaled this cross section to best fit the attachment data for O$_2$, with $F_\text{stat}=1.1\times 10^{-4}$.  These two results are shown in Fig.~\ref{fig:attachment_rates} by the dashed black curves.  

Electron attachment to N$_2$O is dominated by dissociation into NO + O$^-$ at an electron energy of 2.7 eV.  We  have used the cross section of Christophorou (see~\cite{Christophorou}, Fig. 20) plus the direct attachment cross section scaled to 0.006\% of the maximum to compute the attachment rate constant shown by the black dashed line in Fig.~\ref{fig:attachment_rates}.  

The molecules N$_2$, H$_2$O, and CO$_2$ all have electron affinities close to or even less than zero (see Table~\ref{table:properties}, and therefore do not form stable negative ions in the ground state (see page 235 of~\cite{Rienstra-Kiracofe2002}). However, under proper conditions CO$_2$ electron scattering can form a metastable excited complex CO$_2^{*-}$, which auto-dissociates with a lifetime observed to range from 60 to 90 $\mu s$~\cite{Klots1977}.  
The radiative decay of this state is hindered by the Franck-Condon principle because of the difference in shapes of the molecule in the excited and ground states: the bond angle (O--C--O angle) of ground state CO$_2$ is $\sim180^\text{o}$,
while the excited compound state has a bond angle of $\sim135^{\text{o}}$ ~\cite{Krauss1972}.  
This metastable state has been produced in various processes, including electron dissociation of organic molecules containing bent O--C--O bonds~\cite{Cooper1972}, in sputtering from alkali metal surfaces~\cite{Compton2008}, and from (CO$_2$)$_\text {n}$ clusters in supersonic jets~\cite{DeLuca1988}.  It is the metastable state produced by this last process that is presumably responsible for the non-zero attachment rate observed in LAr by Bettini {\it et al.}.  If this is correct, the finite lifetime of the negative ions produced by attachment to CO$_2$ should in principle result in an exponential tail on the charge signal.   With small attachment losses or amplifier shaping times less than several $\mu$s this tail would be unobservable.  However, with larger CO$_2$ concentrations and long shaping times it should be possible to observe the charge "tail" on the electron swarm and confirm the metastability of CO$_2^{*-}$ in LAr.  We have not been able to find a cross section for producing this metastable state, so no calculation for CO$_2$ is shown in Fig.~\ref{fig:attachment_rates}.

The electron affinity of the water molecule has been calculated to be approximately zero~\cite{Chipman1978}.  There is no experimental evidence for the formation of a stable H$_2$O$^-$ ion~\cite{Tsai2004}, but there is a considerable literature on the production of the cluster ions (H$_2$O)$_n^-$~\cite{Lee2005} and (H$_2$O)$_2$Ar$_n^-$~\cite{Tsai2004} by injection of low-energy electrons into the high-pressure region of a seeded supersonic jet.  These stable negative ions all have electron affinities measured to be less than 0.05 eV. If we assume a direct attachment mechanism to form these ions, as above, but scaled by $2.2\times 10^{-6}$ to match the single attachment rate measurement for H$_2$O, we obtain a calculated attachment rate constant shown by the black dashed curve shown in Fig.~\ref{fig:attachment_rates}.  A measurement of attachment to water in gaseous argon~\cite{Wang1984} supports the single low field data point.  The data they show are for the rate constant as a function of electron energy, from 4 eV to 10 eV electron energy.  The electric field at 4 eV from Fig.~\ref{fig:trans_diff} is $\sim$100 kV/cm. The attachment rate indicated at that energy, scaled to the density of LAr, is $3\times10^{10}\text{s}^{-1}$.  At higher energies the rate rises rapidly, which can reasonably be attributed to dissociative attachment.  Clearly more measurements on H$_2$O are needed to determine nature of the attachment process and the field dependence of the attachment rate constant.

If, as suggested, electron attachment to CO$_2$ and H$_2$O only occurs on clusters of two (or more) molecules, then it is necessary to know the concentration of the clusters in the LAr to compute the true attachment rate constant.  However, the concentrations are always measured as single molecules in the gas at room temperature.  If all of the impurity in the LAr is in the form of two-molecule clusters, then the true attachment rate constant would be twice the "apparent" constant computed from the concentration measured in the gas.  In the event that single molecules and molecular clusters were in equilibrium in LAr, the measured rate constant would depend on the impurity concentration.

The nitrogen molecule has an electron affinity of -1.6 eV, and no direct attachment to N$_2$ is possible, and the dissociation energy of N$_2$ is so high that dissociative attachment can only occur for fields well above 100 kV/cm .  It is conceivable that clusters of nitrogen molecules could form and attach electrons, as is assumed for H$_2$O and CO$_2$, but we have been unable to find any reports of this.  Therefore it is reasonable to doubt any reports of attachment to N$_2$.  The large range of values for the data shown in Fig.~\ref{fig:attachment_rates_N2} only makes it more difficult to believe that true attachment to N$_2$ has been observed.  It is possible that what was observed is attachment to oxygen introduced with the nitrogen. The field dependence appears to be consistent with that for O$_2$, as shown in Fig.~\ref{fig:attachment_rates_N2} by the solid black lines which are the best fit attachment rate constant for oxygen scaled by 1.5$\times10^{-3}$,  6.5$\times10^{-4}$, 1.5$\times 10 ^{-4}$, and 2.8$\times^{-7}$ to match the data of Biller {\it et al.}, Hofmann {\it et al.}, Sakai {\it et al.}, and Barabash {\it et al.}, respectively.  Although it might be difficult to justify the large oxygen contaminant required to explain the data of Biller {\it et al.} and of Hofmann {\it et al.}, that would be the simplest explanation of the large values obtained by them.  As an additional cause of the disagreement, Barabash {\it et al.} has proposed a multi-body process involving the participation of the electron, N$_2$, and one or more impurity molecules, other than O$_2$, with positive electron affinities.  Some combination of single and multi-body processes, with reasonable assumptions about N$_2$ and O$_2$ concentrations, might justify the large values and the disagreement among the four data sets.  There is no direct evidence for this.   It would be desirable to make more definitive measurements with several well determined concentrations of both N$_2$ and O$_2$ using a long drift length and nitrogen concentrations higher than Hofmann {\it et al.} and Biller {\it et al.}, but less then those that significantly lower the drift velocity of LAr (i.e. in the range of 10 to 1000 ppm) and very low oxygen concentrations (< 0.1 ppb).  For the present, it seems reasonable to assume that, of the four determinations, attachment to N$_2$ is best represented by the Barabash {\it et al.} data.  That would imply that 0.36\% of N$_2$ is equivalent to 1 ppb of O$_2$.

\section{Discussion}
\begin{figure}[htp]
\centering
\includegraphics[angle=0,width=4.0in]{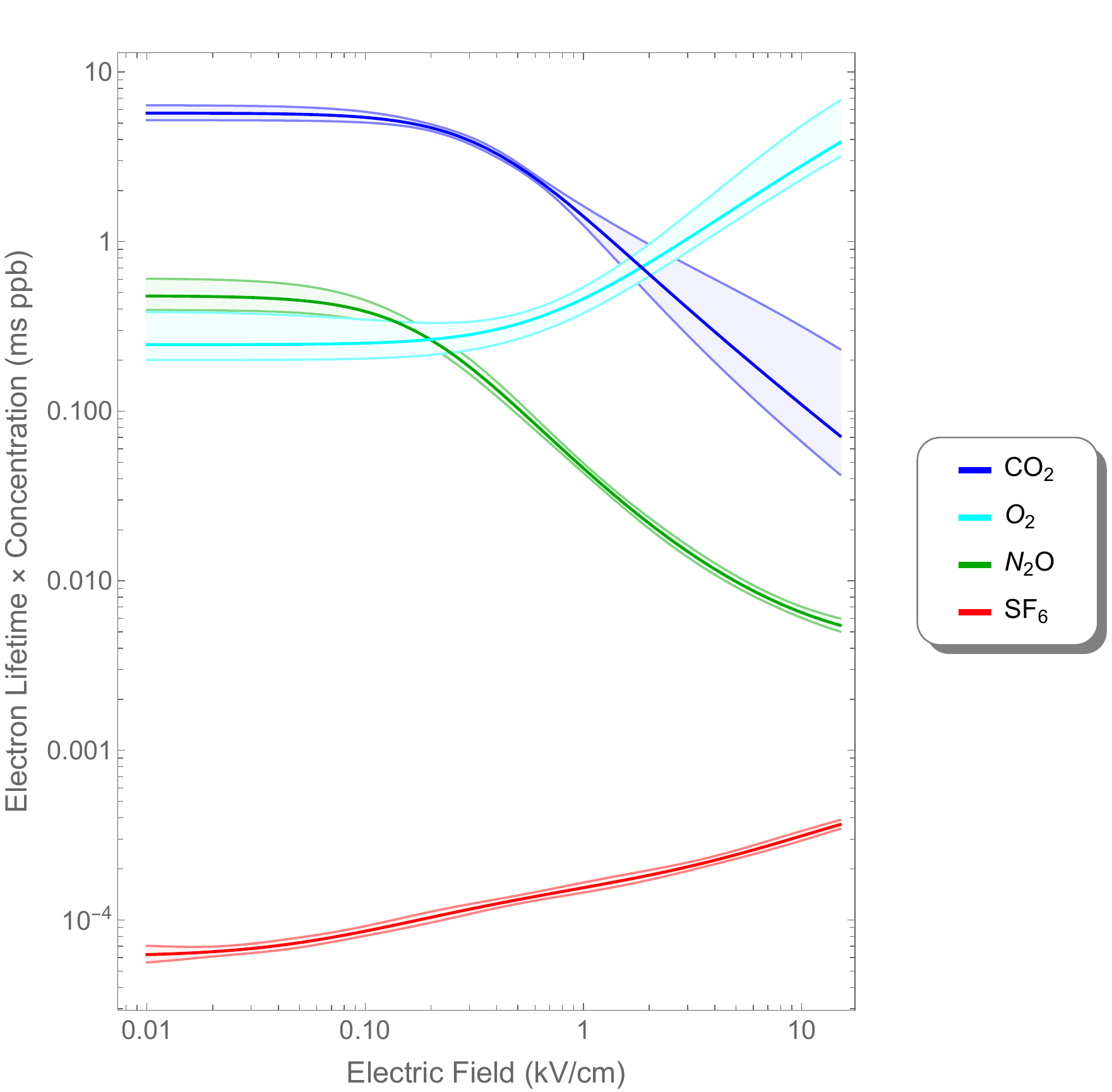}
\caption{Electron lifetime prediction for 1 ppb impurity concentration as a function of electric field for the parameterized attachment rate constants.  To obtain lifetimes at other concentrations divide by the concentration in ppb.}
\label{fig:lifetime}
\end{figure}

It has been demonstrated by ICARUS~\cite{Antonello:2014eha}, MicroBooNE~\cite{microboone2016measurement}, and ProtoDUNE~\cite{DUNE:2020cqd} that very high electron lifetimes ($>$15-30~ms) can be achieved.  Our fits with the parameters shown in Table \ref{table:table_fit} can be used to predict the electron lifetime as a function of electric field for different impurity concentrations at electric fields below a few kV/cm (covering the range of  LArTPC operating conditions), as is shown in Fig.~\ref{fig:lifetime}.
It is clear that a sub-ppb purity level for O$_2$ is required to achieve an electron lifetime of over 1 ms in LAr.  The parameterizations of the attachment coefficients presented here can be particularly useful in addressing the following problems.

\begin{enumerate}
\item If we know the concentrations of one or more impurities, then we can use the attachment rate constant functions to compute the partial lifetime, or attenuation length, for that set of impurities at any electric field. For a mixture, the attachment rates,  computed as  ($k_A \cdot n$) for each impurity, can be summed to produce the total attachment rate.  At low fields, where the attachment rate is independent of the field, the constants listed in Table~\ref{table:table_Zero_Field} can be used instead: for each impurity in column 1, the zero field attachment rates $k_{A,0}$ and attenuation lengths per 1 V/cm, $\tilde{\lambda}_{A,0}$, are given in columns 2 and 3, respectively. These values are correct within the uncertainties stated for fields less than $E_\text{max}$ given in column 4.  This values define the meaning of ``low field'' for each impurity. The remaining columns are discussed below.  The values of $1/k_{A,0}$, when divided by the impurity concentration in ppb, gives the lifetime $\tau$ in ms; the factor $\tilde{\lambda}_{A,0}$, when multiplied by the field in V/cm and divided by the concentration in ppb, gives the attenuation length $\lambda$ in cm.

\begin{table}[!htbp]
\caption{Quantities characterizing attachment at zero field.}
\label{table:table_Zero_Field}
\begin{tabular}{|c|c|c|c|c|c|}\hline

$\text{Molecule}$ & $k_\text{A,0}\text{ (ms ppb)}^{-1}$  & $ \tilde \lambda_{A,0} \text{ cm}^2 \text{ V}^{-1} \text{ppb)}$  & $ E_\text{max}\text{ (V/cm)}$  & $ k_\text{A,500}/k_\text{A,0}$  & $ E_\text{min}\text{ (V/cm)}$  \\
\hline
 $\text{SF}_6$  & $ (1.64\pm0.29)\times10^{4}$  & $0.00003(4\pm7) $ &  21 & $0.46\pm0.09$  & 389 \\
 $ \text{N}_2\text{O}$  & $ 2.1\pm0.4$  & $ 0.26\pm 0.07$  & 94 & $ 4.6\pm1.1$  & 462 \\
 $ \text{O}_2$  & $ 4.1_{\scriptscriptstyle-0.9}^{\scriptscriptstyle+1.5}$  & $ 0.14_{\scriptscriptstyle-0.04}^{\scriptscriptstyle+0.05}$  & 315 & $ 0.75_{\scriptscriptstyle-0.25}^{\scriptscriptstyle+0.29}$  & 86 \\
 $ \text{CO}_2$  & $ 0.175\pm0.018$  & $ 3.2\pm0.5$  & 131 & $ 2.06\pm0.23$  & 473 \\
 $ \text{H}_2\text{O}$  & $ 0.093_{\scriptscriptstyle-0.033}^{\scriptscriptstyle+0.028}$  & $ 5.9_{\scriptscriptstyle-2.3}^{\scriptscriptstyle+1.9}$  & $ \text{unknown}$  & $ \text{unknown}$  & $ \text{unknown}$  \\
\hline
\end{tabular}
\end{table}

\item We can convert measured lifetimes to an "O$_2$ equivalent" concentration or to any other "impurity equivalent", or given a set of concentrations (as for example air) we can compute the "impurity equivalent" concentration for that mixture. 

\item We observe that the log of the zero field attachment for the four molecules with positive electron affinities is roughly proportional to the electron affinity (within a factor of 2.5 higher or lower): 
\begin{equation}
 Log[k_A(0) (s^{-1})] = 24.9 +11.45 E_A \text{(eV)}.
\label{eq:log_kA-est}
\end{equation}
This offers the possibility of estimating the impact on electron lifetime for other impurities, whose electron affinities are known.  Reference ~\cite{Rienstra-Kiracofe2002} contains a table of electron affinities for more than 1000 atoms and molecules.

\item For large attenuation lengths, the charge attenuation as described in  Eq.~\eqref{eq:att} is small.  The precision of the charge attenuation determination is then limited by electronic noise, and systematic uncertainties in grid transmission and signal processing (see ~\cite{microboone2016measurement}).  To improve the precision of charge attenuation determination, one strategy is to measure the attenuation at a lower electric field, so that the attenuation can be significantly increased and more accurately measured. It is straightforward to show that if the purity monitor operation is described simply by Eq.~\ref{eq:att}, and if the noise in the initial (cathode) and final (anode) charge signals is the same, then the minimum uncertainty in the determination of the lifetime occurs when the lifetime is slightly less than one half of the drift time. Reducing the field to achieve this condition is limited by the fact that when the field is too low the longitudinal diffusion time of the swarm will exceed the shaping time of the charge amplifier.  This will result in a systematic reduction in the measured charge at the cathode and an erroneously low value of attenuation length.  In addition, at very low fields, the increase of the effect of diffusion on grid transparency will also lead to systematically low values of the attenuation length. These two systematic effects can be minimized and corrected for by appropriate design of the instrumentation and by modeling of the effects.  Finally, the reduction in cathode quantum efficiency at low fields (see Fig. 9 of~\cite{Li:2015rqa}) will result in an increased statistical uncertainty in charge measurement and therefore in attenuation length determination.

\item When we have measured the attenuation length at a lower field to obtain the most accurate measurement, we must convert the result to the attenuation length at the operating field of the LArTPC. This requires an accurate conversion factor to convert from the value at the low electric field setting to that at the nominal electric field setting. This can, of course be done by evaluating the parameterizations of Table~\ref{table:table_Zero_Field} at the two fields.  For the typical operating field of 500 V/cm, the conversion factors $k_{A,500}/k_{A,0}$ in column 5 of Table~\ref{table:table_Zero_Field} can be used. These factors are only applicable if the low field is less than the maximum field E$_\text{max}$ given in column 4. For fields greater than the field given in column 6 but less than 500 V/cm, the attachment rates at the two fields are the same within the uncertainties.  In any case, since the factor depends on the particular mix of impurities, to be correct the conversion factor requires a priori knowledge of the amounts of all the impurities in the LAr.

\item When we know the concentrations of a set of impurities, we can compute the partial lifetime due to these and compare it to the total lifetime measured by the purity monitor to determine if there is any evidence for impurities other than those whose concentrations we have measured. The missing amount of the unknown impurity can be expressed as an O$_2$-equivalent concentration, or simply as an unassigned partial lifetime.
\end{enumerate}

\section{Conclusions}~\label{conclusions}
In this paper we summarize data obtained from the literature for the electron attachment rate constants as a function of the external electric field for SF$_6$, O$_2$, N$_2$O, CO$_2$, H$_2$O, and N$_2$ in LAr. We further provide rational polynomial functions of electric field that best represent -- in the least-squares sense -- this data. These functions can be useful when comparing measurements to modeling and in analyzing detector performance.  We show uncertainty bands for these "best-fit" functions and provide $\chi^2/\text{DOF}$ values indicating that they are good representations of the data, given the statistical uncertainties.  For oxygen we show that the five available data sets have systematic uncertainties larger than the stated statistical uncertainties, and estimate the total uncertainty in our 
"best-fit" function.  Examples of applying these parameterizations to make interpolations and understand performance are presented. It is furthermore shown that a very simple calculation can reasonably describe the major features of the energy dependence of the attachment rate constants for different molecules.  We have presented evidence that electron attachment to nitrogen in LAr is extremely small compared to oxygen, but not quantitatively understood.  Attachment to water is defined by a single measurement at a very low field and further measurements are required to verify this single low field point and to extend the results to the higher fields of present detectors.  Most of the measurements discussed here were made more than 25 years ago, before the availability of commercial instruments capable of  measuring impurity concentrations in argon gas with a few percent accuracy over a large dynamic range ($\sim$10$^2$ ppt to many ppm), before the development and refinement of the ICARUS purity monitor, and before the refinement of the materials and technology for purifying LAr. As a consequence, it is now possible to measure electron attachment rate constants in LAr with much higher accuracy than that of the data reviewed here.  In fact, the instrumentation available now is so superior to that used in the past that a program of attachment measurements should be made to resolve the many ambiguities, uncertainties, and unknowns -- some of which have been discussed here.  
 Finally, we would suggest the importance of further measurements to determine whether impurities other than oxygen and water significantly contribute to the finite electron lifetimes observed in the present LArTPCs, and to estimate the practical upper limit on the drift length in LAr imposed by attachment.

\acknowledgments
This work is supported by Brookhaven National Laboratory and U.S. Department of Energy, Office of Science, 
Office of High Energy Physics under contract number DE-SC0012704.
\bibliographystyle{hunsrt}
\bibliography{sum_att}{}

\newpage
\section*{Appendix: Extracted data from the literature}
\small
\begin{minipage}[t]{.5\textwidth}
Adams {\it et al.} data \cite{adams2005purity} for O$_2$:
\begin{centermath}
\begin{array}{c|c}
\text{E (kV/cm)} &  k_A \pm  1\sigma  (s^{-1}) \\
\hline
 2.20 & (9.3\pm0.7)\times10^{11} \\
 2.40 & (8.6\pm0.6)\times10^{11} \\
 4.40 & (6.2\pm0.4)\times10^{11} \\
 4.80 & (5.2\pm0.4)\times10^{11} \\
 6.20 & (4.5\pm0.6)\times10^{11} \\
\end{array}
\end{centermath}
Biller {\it et al.} data \cite{Biller:1989yq}  for N$_2$:
\begin{centermath}
\begin{array}{c|c}
\text{E (kV/cm)} &  k_A \pm  1\sigma  (s^{-1}) \\
\hline
 0.50 & (3.4\pm0.5)\times10^9 \\
 1.00 & (3.22\pm0.25)\times10^9 \\
 1.50 & (2.74\pm0.18)\times10^9 \\
 2.00 & (2.47\pm0.20)\times10^9 \\
 3.00 & (1.77\pm0.11)\times10^9 \\
\end{array}
\end{centermath}
Biller {\it et al.} data \cite{Biller:1989yq} for O$_2$:
\begin{centermath}
\begin{array}{c|c}
\text{E (kV/cm)} & k_A \pm  1\sigma  (s^{-1}) \\
\hline
0.50 & (2.79\pm0.20)\times10^{12} \\
1.00 & (2.22\pm0.13)\times10^{12} \\
1.50 & (1.85\pm0.08)\times10^{12} \\
2.00 & (1.56\pm0.05)\times10^{12} \\
3.00 & (1.04\pm0.06)\times10^{12} \\
\end{array}
\end{centermath}

Bakale {\it et al.} data\cite{bakale1976effect} for N$_2$O (no error provided in the original figure; assigned $\delta k_A/k_A = 20.2\%$):
\begin{centermath}
\begin{array}{c|c}
\text{E (kV/cm)} & k_A (s^{-1}) \\
\hline
0.07 & 3.83\times10^{12} \\
0.10 & 2.09\times10^{12} \\
0.25 & 4.91\times10^{12} \\
0.51 & 9.41\times10^{12} \\
0.50 & 1.10\times10^{13} \\
0.75 & 1.44\times10^{13} \\
0.76 & 1.27\times10^{13} \\
1.01 & 2.16\times10^{13} \\
1.03 & 2.73\times10^{13} \\
1.25 & 3.31\times10^{13} \\
1.52 & 4.00\times10^{13} \\
1.53 & 4.79\times10^{13} \\
2.63 & 5.67\times10^{13} \\
4.29 & 7.25\times10^{13} \\
5.10 & 8.78\times10^{13} \\
6.26 & 1.04\times10^{14} \\
7.60 & 1.61\times10^{14} \\
10.35 & 2.25\times10^{14} \\
15.26 & 2.70\times10^{14} \\
25.52 & 2.33\times10^{14} \\
40.78 & 2.04\times10^{14} \\
51.25 & 2.28\times10^{14} \\
71.38 & 1.86\times10^{14} \\
80.03 & 1.95\times10^{14} \\
88.69 & 2.18\times10^{14} \\
104.10 & 2.38\times10^{14} \\
\end{array}
\end{centermath}
\end{minipage}
\begin{minipage}[t]{.5\textwidth}
  Bakale {\it et al.} data\cite{bakale1976effect} for O$_2$ (no error provided in the orginal figure; assigned $\delta k_A/k_A = 6.9\%$):
\begin{centermath}
\begin{array}{c|c}
\text{E (kV/cm)} &  k_A (s^{-1})\\
\hline
0.05 & 5.87\times10^{12} \\
0.10 & 4.80\times10^{12} \\
0.15 & 4.49\times10^{12} \\
0.20 & 4.49\times10^{12} \\
0.39 & 4.19\times10^{12} \\
0.50 & 3.79\times10^{12} \\
0.73 & 2.62\times10^{12} \\
0.97 & 2.42\times10^{12} \\
1.45 & 2.26\times10^{12} \\
2.91 & 1.65\times10^{12} \\
4.92 & 7.88\times10^{11} \\
7.26 & 6.73\times10^{11} \\
9.66 & 6.73\times10^{11} \\
\end{array}
\end{centermath}

Bakale {\it et al.} data \cite{bakale1976effect} for SF$_6$ (no error provided in the original figure; assigned $\delta k_A/k_A = 6.2\%$):
\begin{centermath}\begin{array}{c|c}
\text{E (kV/cm)} & k_A (s^{-1}) \\
\hline
0.01 & 1.55\times10^{16} \\
0.02 & 1.60\times10^{16} \\
0.03 & 1.58\times10^{16} \\
0.04 & 1.29\times10^{16} \\
0.05 & 1.46\times10^{16} \\
0.07 & 1.24\times10^{16} \\
0.10 & 1.16\times10^{16} \\
0.21 & 9.35\times10^{15} \\
0.31 & 8.54\times10^{15} \\
0.42 & 8.84\times10^{15} \\
0.51 & 7.47\times10^{15} \\
0.73 & 6.75\times10^{15} \\
1.04 & 6.10\times10^{15} \\
2.07 & 5.90\times10^{15} \\
3.19 & 4.93\times10^{15} \\
4.15 & 4.12\times10^{15} \\
5.27 & 4.07\times10^{15} \\
7.18 & 3.37\times10^{15} \\
10.35 & 3.25\times10^{15} \\
15.62 & 2.94\times10^{15} \\
20.54 & 2.46\times10^{15} \\
31.00 & 2.05\times10^{15} \\
\end{array}
\end{centermath}
\end{minipage}
\begin{minipage}[t]{.5\textwidth}
Bettini {\it et al.} data \cite{bettini1991study} for O$_2$:
\begin{centermath}
\begin{array}{c|c}
\text{E (kV/cm)} &  k_A \pm  1\sigma  (s^{-1}) \\
\hline
0.10 & (3.52\pm0.25)\times10^{12} \\
0.20 & (3.56\pm0.26)\times10^{12} \\
0.30 & (3.47\pm0.25)\times10^{12} \\
0.50 & (3.26\pm0.24)\times10^{12} \\
0.60 & (3.09\pm0.23)\times10^{12} \\
0.70 & (2.99\pm0.24)\times10^{12} \\
0.80 & (2.78\pm0.21)\times10^{12} \\
\end{array}
\end{centermath}

Bettini {\it et al.} data \cite{bettini1991study} for CO$_2$:
\begin{centermath}
\begin{array}{c|c}
\text{E (kV/cm)} &  k_A \pm  1\sigma  (s^{-1}) \\
\hline
0.10 & (1.86\pm0.14)\times10^{11} \\
0.20 & (2.12\pm0.16)\times10^{11} \\
0.30 & (2.53\pm0.19)\times10^{11} \\
0.40 & (3.02\pm0.25)\times10^{11} \\
0.50 & (3.62\pm0.30)\times10^{11} \\
0.60 & (4.2\pm0.4)\times10^{11} \\
0.70 & (4.9\pm0.5)\times10^{11} \\
0.80 & (5.6\pm0.5)\times10^{11} \\
\end{array}
\end{centermath}

Carls, {\it et al.} data\cite{microboone2016measurement} for H$_2$O
\begin{centermath}
\begin{array}{c|c}
\text{E (kV/cm)} &  k_A \pm  1\sigma  (s^{-1}) \\
\hline
0.03 & 9.3_{-3.3}^{+2.8}\times10^{10}  \\
\end{array}
\end{centermath}
Hofmann {\it et al.} data \cite{hofmann1976production} for O$_2$:
\begin{centermath}
\begin{array}{c|c}
\text{E (kV/cm)} &  k_A \pm  1\sigma  (s^{-1}) \\
\hline
2.01 & (1.7\pm0.4)\times10^{12} \\
2.51 & (1.51\pm0.27)\times10^{12} \\
3.01 & (1.18\pm0.13)\times10^{12} \\
3.01 & (8.9\pm1.0)\times10^{11} \\
4.03 & (6.4\pm1.4)\times10^{11} \\
5.55 & (5.2\pm1.0)\times10^{11} \\
6.56 & (4.4\pm1.1)\times10^{11} \\
7.49 & (3.4\pm0.9)\times10^{11} \\
8.98 & (3.2\pm0.9)\times10^{11} \\
11.72 & (2.53\pm0.21)\times10^{11} \\
19.95 & (2.11\pm0.06)\times10^{11} \\
19.98 & (1.88\pm0.09)\times10^{11} \\
24.94 & (1.81\pm0.09)\times10^{11} \\
26.20 & (1.57\pm0.07)\times10^{11} \\
\end{array}
\end{centermath}
\end{minipage}
\begin{minipage}[t]{.5\textwidth}
Hofmann {\it et al.} data \cite{hofmann1976production} for N$_2$:
\begin{centermath}
\begin{array}{c|c}
\text{E (kV/cm)} &  k_A \pm  1\sigma  (s^{-1}) \\
\hline
2.0 & (5.2\pm 2.3)\times10^8 \\
5.0 & (5.8\pm 2.1)\times10^8 \\
\end{array}
\end{centermath}
Barabash {\it et al.} data \cite{barabash1984} for N$_2$:
\begin{centermath}
\begin{array}{c|c}
\text{E (kV/cm)} &  k_A \pm  1\sigma  (s^{-1}) \\
\hline
0.43 & (5.\pm4.)\times10^5 \\
0.88 & (5.7\pm3.3\times10^5 \\
1.34 & (5.4\pm2.0)\times10^5 \\
1.78 & (4.7\pm1.3)\times10^5 \\
2.23 & (3.9\pm1.8)\times10^5 \\
2.67 & (3.1\pm2.6)\times10^5 \\
\end{array}
\end{centermath}

Sakai {\it et al.} data \cite{Sakai1984} for N$_2$ (no error provided in the original figure; assigned $\delta k_A/k_A = 37.2\%$
\begin{centermath}
\begin{array}{c|c}
\text{E (kV/cm)} &  k_A \pm  1\sigma  (s^{-1}) \\
\hline
 4.9 & (6.8\pm 2.5)\times 10^7 \\
 7.4 & (9.0\pm 3.3)\times 10^7 \\
 9.9 & (3.7\pm 1.4)\times 10^7 \\
 12.5 & (3.7\pm 1.4)\times 10^7 \\
 12.4 & (2.9\pm 1.1)\times 10^7 \\
 15.0 & (2.2\pm 0.8)\times 10^7 \\
 15.0 & (1.8\pm 0.7)\times 10^7 \\
 4.7 & (1.2\pm 0.4)\times 10^8 \\
 7.2 & (8.9\pm 3.3)\times 10^7 \\
 12.0 & (5.5\pm 2.1)\times 10^7 \\
 14.5 & (4.7\pm 1.8)\times 10^7 \\
 16.9 & (3.7\pm 1.4)\times 10^7 \\
 21.5 & (2.8\pm 1.0)\times 10^7 \\
 23.8 & (2.8\pm 1.0)\times 10^7 \\
 26.2 & (2.8\pm 1.0)\times 10^7 \\
 28.6 & (2.4\pm 0.9)\times 10^7 \\
 33.5 & (2.1\pm 0.8)\times 10^7 \\
 38.4 & (1.3\pm 0.5)\times 10^7 \\
 10.0 & (4.9\pm 1.8)\times 10^7 \\
 12.5 & (4.8\pm 1.8)\times 10^7 \\
 15.0 & (4.5\pm 1.7)\times 10^7 \\
 20.0 & (3.9\pm 1.4)\times 10^7 \\
 24.8 & (3.7\pm 1.4)\times 10^7 \\
 29.8 & (2.9\pm 1.1)\times 10^7 \\
 34.8 & (3.0\pm 1.1)\times 10^7 \\
 40. & (2.7\pm 1.0)\times 10^7 \\
\end{array}
\end{centermath}
\end{minipage}

\end{document}